# Optical properties of single metallic nanorods: An analytical model

Cheng-ping Huang[1]* and Xiao-gang Yin[2]

[1]*Department of Physics, Nanjing Tech University, Nanjing 211816, China*

[2]*College of Physics, Nanjing University of Aeronautics and Astronautics, Nanjing 211106, China*

## Abstract

It is well known that the optical properties of spherical metal particles can be described with the Rayleigh approximation or rigorous Mie theory. But for the single metallic nanorods, a theory well capturing the fundamental resonance and scattering features is still absent. In this study, an analytical model is developed for the metallic nanorod, considering the longitudinally non-uniform conduction current and surface charges. With the circuit parameters deduced from the kinetic and electromagnetic energy of the nanorod, a formula which agrees well with the simulations has been suggested for the resonance wavelength of the fundamental longitudinal mode. Moreover, by introducing the radiative resistance to the circuit theory, the dipole moment, extinction spectrum, and near-field enhancement of the nanorod have been derived analytically and confirmed numerically. The results are important for understanding the optical properties of the metallic nanorods and provide a guideline for designing the light scattering and absorption.

* Email: cphuang@njtech.edu.cn

# I. Introduction

The photons and electrons, which are of different nature, may be coupled to generate the quasi-particles, i.e., the plasmon polaritons. For a flat metal surface, surface-plasmon polariton can be excited, propagating along the surface with subwavelength mode sizes. This surface mode plays an important role in the plasmonic waveguiding and light transmission through perforated metal films [1, 2]. On the other hand, the small metal particles, such as the nanospheres and nanorods etc., can support the localized surface-plasmon (LSP) resonance. The LSP resonance gives rise to strong field enhancement near the particle and enhanced scattering and absorption of incident light. The localized excitation has a number of potential applications, such as the plasmonic sensing [3], enhancement of Raman signal and harmonic generation [4, 5], photo-thermal therapy [6], and information storage [7], etc. Moreover, by using arrays of metal particles, various optical structures or devices such as the plasmonic waveguides [8, 9], directional optical antennas [10], ultrathin metasurfaces [11-13], and plasmonic crystals [14-16] can be constructed.

Besides the experimental progress achieved so far, the theoretical studies are especially important for understanding the resonance effect and designing the particle-based devices. As is well known, for the spherical metal particles, the resonance effect can be uncovered with the Rayleigh approximation or Mie scattering theory [17]. Compared with the metal spheres, the fundamental resonance mode along the longitudinal direction of the nanorods is strongly dependent on the aspect ratio [18], providing an efficient way to tune the plasmonic resonance and thus attracting much more interest. But, up to now, a theory well capturing the fundamental resonance and scattering features of a metallic nanorod is still absent. By modeling the cylindrical nanorod as elongated ellipsoid, Gans theory and empirical formulas with fitting parameters have been suggested to describe the resonance wavelength or scattering intensity [19-21]. However, the numerical simulations demonstrated that the plasmonic resonance features of the cylindrical nanorod differ significantly from the ellipsoid [22]. Starting from the waveguide theory, Novotny has derived successfully the effective wavelength of an optical antenna consisting of a single metallic nanorod [23]. Nonetheless, the complicated format makes it difficult to arrive at a direct and accurate expression for the resonance wavelength, especially when the rod radius is not too small ($r_0>5$ nm). Moreover, the scattering and absorption features of the nanorod have not been addressed by the waveguide theory. In 2009, the circuit theory was employed by the present authors and coauthors to study the fundamental plasmonic resonance of the nanorod [24]. But some excessive approximations or treatments, such as the homogeneous current distribution along the nanorod, the capacitance of nanorod with a fitting parameter [25] etc., have been used. More importantly, the width and magnitude of the scattering, absorption, and extinction spectrum predicted by the theory do not agree with the numerical simulations.

In this paper, an analytical model has been developed for the metallic nanorod, which well describes the fundamental plasmonic resonance wavelength, the scattering /absorption/extinction spectrum, and the near-field enhancement, etc. The model is based on the circuit theory but the theoretical analysis highlights the energy aspect of the nanorod. In the theoretical treatment, non-uniform distributions of the conduction currents and surface charges along the nanorod have been considered and the radiative resistance has been included in the modified circuit equation. These treatments contribute to the good agreement between the theory and simulations and are helpful for uncovering the underlying physics. The paper is organized as follows. In Sec. II, the non-uniform distributions of conduction current and surface charges along the nanorod have been discussed. An analytical formula for the plasmonic resonance wavelength of the nanorod is obtained in Sec. III. The dependence of resonance wavelength on the geometrical and material parameters has been suggested and compared with the simulations. In Sec. IV, the dipole moment of the nanorod is deduced with the circuit equation, where the radiative resistance has been determined and employed. The resonance broadening associated with both the radiative resistance and Faraday inductance is revealed. The scattering, absorption, and extinction cross sections are given in Sec. V. The evolution of extinction spectrum with the parameters is discussed and a systematic comparison between the theory and simulations is given as well. The Sec. VI calculates the enhancement ratios of the electric and magnetic field in the near zone of the nanorod. And a short summary is presented in Sec. VII.

## II. Current & charge distributions

The metallic nanorod considered here is schematically shown in Fig. 1(a), which usually has two semispherical ends as demonstrated by the experiments [18, 19]. The radius of nanorod is $r_0$ and the total length is $l$ ($l \gg r_0$). The nanorod is embedded in a dielectric medium with the permittivity of $\varepsilon_d$. The incident light propagates with the electric field along the axis of nanorod, exciting the longitudinal plasmonic resonance (here we are interested in the fundamental mode). We suppose that the radius of the nanorod is much larger than the Fermi wavelength but smaller than the skin depth of metal [$\lambda_F (\sim 0.5\ nm) \ll r_0 \leq \delta\ (\sim 20\ nm)$], thus the quantum effect can be neglected and the fields in the cross section of the nanorod can be taken to be homogeneous.

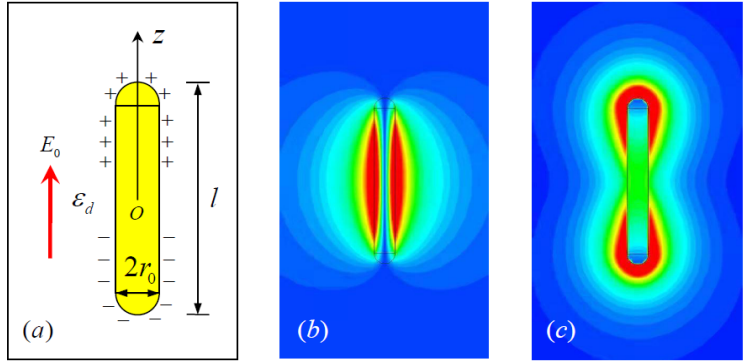

*Fig. 1 (a) Schematic view of single metallic nanorod, which is embedded in a dielectric medium and excited by a longitudinal light electric field. (b, c) Amplitude distributions of magnetic field (b) and electric field (c) at the plasmonic resonance.*

In the optical frequency range, the free electrons in the metallic nanorod play a dominant role and the effect of bounded electrons can be ignored. Under the action of longitudinal light electric field, an oscillating conduction current can be induced in the nanorod. Correspondingly, positive and negative charges are accumulated on the nanorod surface, i.e., the side and ends of the nanorod. The induced current and charges can excite a strong magnetic field [Fig. 1(b)] and electric field [Fig. 1(c)] around the metal nanorod. Numerical simulation showed that, for the fundamental longitudinal mode, the conduction current along the nanorod is not uniformly distributed. Instead, a standing-wave-like pattern is present for the conduction current, which is maximal at the center of nanorod and zero near the two ends. For a longer nanorod, the complex conduction current density of the fundamental mode can be approximated by [23]

$$J(z) = J_0 \cos(\pi z / l) e^{-i\omega t}, \tag{1}$$

where $z \in [-l/2, l/2]$ and $J_0$ is the magnitude of current density at the center of nanorod (this approximation will be failed when the nanorod length is comparable with the diameter). Correspondingly, the surface density of free charges can be expressed as

$$\rho(z) = i\rho_0 \sin(\pi z / l) e^{-i\omega t}, \tag{2}$$

where $\rho_0$ is the magnitude of surface charge density near the two ends. By using the law of charge conservation $SdJ = -d(\rho 2\pi r_0 dz)/dt$ in the volume element $dV = Sdz$ ($S = \pi r_0^2$ is the cross-sectional area of the nanorod), the relationship between the magnitude of surface charges and conduction current can be determined as

$$J_0 = (2l / \pi r_0) \rho_0 \omega. \tag{3}$$

With the Eq. (2), some useful results can be deduced. One result is that the total surface charges accumulated on the upper half part of the nanorod are

$$Q \approx \rho(0.5l) 2\pi r_0^2 + \int_0^{0.5l - r_0} \rho(z) 2\pi r_0 dz \tag{4}$$
$$= iQ_0 e^{-i\omega t}.$$

Here, the first and second term on the right-hand side of Eq. (4) represents the charges accumulated on the end and side of the nanorod, respectively (the surface charge density on the end can be supposed to be uniform for the long nanorod); $Q_0 = 2\rho_0 r_0 l$ is the magnitude of oscillating free charges on half a nanorod. We can deduce that the fraction of charges distributed on the semispherical end is $R = \pi r_0 / l$. For a nanorod with an aspect ratio of $\kappa = l/2r_0 = 5$, for example, about 31% surface charges distribute on the end and 69% charges on the side of nanorod.

Another result is the dipole moment of the nanorod. For convenience, we treat all the positive (or negative) free charges as a point charge $Q$ (or $-Q$) with the position locating at the center of charges $z_0$ (or $-z_0$). With the Eqs. (2) and (4), we obtain

$$z_0 = \int z dq / Q = l / \pi. \tag{5}$$

Then, the distance between the centers of positive and negative charges is $\Delta z = 2l/\pi$. Correspondingly, the dipole moment of the nanorod is

$$p = (2/\pi) Q l. \tag{6}$$

Compared with the case that the charges distribute completely on the ends of nanorod [24], the dipole moment is reduced by a factor of $2/\pi$.

### III. Resonance wavelength

*3.1 Analytical formula of resonance wavelength*

If the sizes of the metal particles are much smaller than the wavelength, the quasi-static approach may be applicable and the circuit parameters such as the inductance and capacitance can be used [26]. When the metal nanorod is excited, there is kinetic energy of moving free electrons, electric-field energy of surface charges, and magnetic-field energy of conduction current stored in or near the nanorod. Consequently, by calculating the kinetic and electromagnetic energy analytically, the circuit parameters of the nanorod may be determined [27].

Based on the above results, the kinetic and electromagnetic energy of free charges or currents have been calculated and the circuit parameters have been derived (see Appendix A-C). For the kinetic energy of free electrons, by setting $U_K = (1/2) L_K I_m^2$ (where $I_m = J_0 S \cos \omega t$ is the real and spatially maximal current at the center of nanorod), the kinetic inductance is obtained as (Appendix A):

$$L_K = \frac{\mu_0 l}{2\pi}(\frac{\delta}{r_0})^2. \tag{7}$$

Here, $\delta$ is the skin depth of the metal. Thus, the kinetic inductance is dominated by the length and radius of the nanorod. Especially, it will play a critical role when the rod radius is comparable with or smaller than the skin depth of the metal. Due to the decrease of electron velocity near the ends of nanorod, the kinetic inductance is significantly reduced, which is just half the result of the uniform-current case [24].

The conduction current will excite an enhanced magnetic field within and outside the nanorod. By calculating the total magnetic-field energy approximately and setting $U_M = (1/2)L_F I_m^2$, the Faraday inductance of the nanorod can be obtained as follows (Appendix B):

$$L_F = \frac{\mu_0 l}{4\pi}(\ln 2\kappa + \frac{1}{4}). \tag{8}$$

Thus, the Faraday inductance is governed by the length $l$ and aspect ratio $\kappa$ of the nanorod. Similarly, compared with the case of longitudinally uniform current distribution, the Faraday inductance is also reduced.

The electric energy includes not only the electric-field energy of the isolated positive and negative charges but also the interacting electric potential energy of them. By setting $U_E = Q_m^2/2C$ (where $Q_m = Q_0 \sin \omega t$), the capacitance of nanorod is determined as (Appendix C):

$$C = \frac{4\varepsilon_0 \varepsilon_d \kappa l}{(2\pi - \kappa) + \pi(\kappa - 2)\ln \kappa}. \tag{9}$$

Thus, the capacitance of nanorod is dependent on the rod length $l$, aspect ratio $\kappa$, and the permittivity $\varepsilon_d$ of the surrounding medium.

At the free resonance (the Ohmic loss and radiation loss of the nanorod are neglected), the total kinetic and electromagnetic energy of the system will be conserved. That is

$$\frac{1}{2}L_K I_m^2 + \frac{1}{2}L_F I_m^2 + \frac{Q_m^2}{2C} = const. \tag{10}$$

By calculating the derivative of the total energy versus time and noticing that $I_m = dQ_m/dt$, one obtains

$$\frac{d^2}{dt^2}Q_m + \frac{1}{(L_K + L_F)C}Q_m = 0. \tag{11}$$

Thus the resonance frequency of the fundamental longitudinal mode of the metallic nanorod is

$$\omega_0 = \frac{1}{\sqrt{(L_K + L_F)C}}. \tag{12}$$

By substituting the kinetic inductance, Faraday inductance, and capacitance into the Eq. (12), the vacuum resonance wavelength can be derived as

$$\lambda_0 = 4\kappa n_d \sqrt{\frac{2\delta^2 + (\ln 2\kappa + 1/4)r_0^2}{(2/\kappa - 1/\pi) + (1 - 2/\kappa)\ln \kappa}}. \tag{13}$$

Here, $n_d$ is the refractive index of the surrounding medium. Thus, the resonance wavelength is determined by the material properties (the surrounding medium index $n_d$ and the metal skin depth $\delta$) and the geometrical parameters (the aspect ratio $\kappa$ and the rod radius $r_0$) of the nanorod. Especially, the resonance wavelength is proportional to the refractive index $n_d$, which provides an efficient tuning method.

*3.2 Dependence of resonance wavelength on the parameters*

Based on the Eq. (13), the dependence of resonance wavelength on the aspect ratio, rod radius, and medium index has been calculated and the results are plotted with the lines in Figs. 2(a), 2(b) and 2(c), respectively. Moreover, to verify the validity of theory, numerical simulations based on the finite-difference time-domain (FDTD) method have been carried out [28]. In the simulation, the dispersion of metal is described with the Drude model $\varepsilon_m = 1 - \omega_p^2/\omega(\omega + i\gamma)$, where the plasma frequency is $\omega_p = 1.37 \times 10^{16}$ rad/s (corresponding to a skin depth $\delta \approx 22\ nm$) and the collision frequency of free electrons is $\gamma = 5 \times 10^{13}\ Hz$. The mesh size in the nanorod is set as 0.2-1.0 nm, depending on the sizes of the nanorod (a larger mesh size is used for the larger nanorod). With the numerical simulations, the extinction spectrum of the nanorod can be obtained (as shown in Sec. V) and the resonance wavelength extracted from the extinction spectra is mapped in Fig. 2 by the circles.

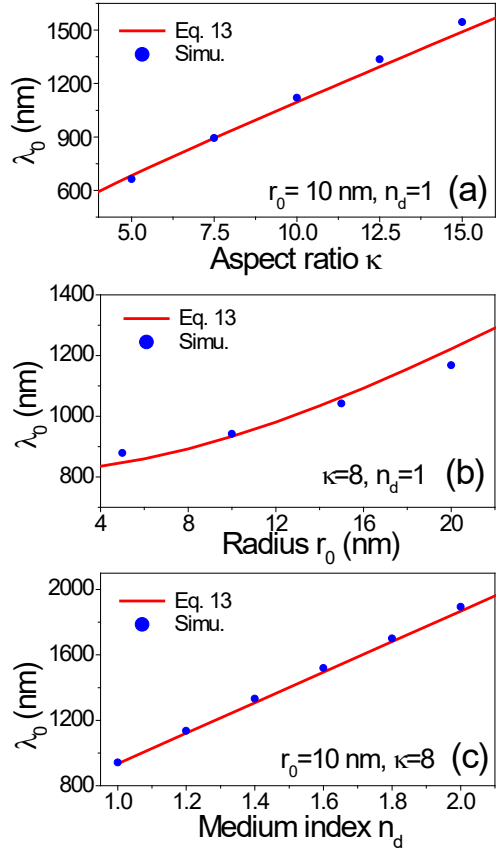

*Fig. 2 Resonance wavelength of the nanorod as a function of (a) aspect ratio ($r_0=10$ nm, $n_d=1$), (b) the rod radius ($\kappa=8$, $n_d=1$), and (c) the medium index ($\kappa=8$, $r_0=10$ nm), respectively. Here, the lines represent the analytical results and the circles the simulation results.*

The resonance wavelength as a function of aspect ratio is shown in Fig. 2(a), where the nanorod is in vacuum and the rod radius is fixed as 10 nm. The resonance wavelength increases almost linearly with the aspect ratio $\kappa$. When $\kappa$ varies from 5 to 15 (or $l$=100-300 nm), the resonance wavelength predicted by the Eq. (13) increases from 683 to 1490 nm, approaching the simulation values from 663 to 1544 nm (the largest deviation is 3.5% for $\kappa=15$). In Fig. 2(b), the resonance wavelength as a function of rod radius is shown, where the nanorod with a constant aspect ratio $\kappa=8$ is placed in vacuum. One can see that the resonance wavelength grows obviously and nonlinearly with the rod radius. When $r_0$ varies from 5 to 20 nm, the resonance wavelength increases from 846 to 1221 nm analytically and from 879 to 1168 nm numerically (the maximal deviation is 4.5% for $r_0$=20 nm). Thus the aspect ratio alone cannot determine the plasmonic resonance wavelength [22]. This is associated with the Faraday inductance [according to the Eq. (12), the resonance wavelength will be independent of the rod radius if $L_F$ is neglected]. In addition, we fix the sizes of nanorod ($\kappa$=8 and $r_0$=10 nm) and change the refractive index of the surrounding medium. The resonance wavelength as a function of medium index is present in Fig. 2(c). When $n_d$ changes from 1.0 to 2.0, the analytical resonance wavelength increases linearly from 934 to 1868 nm, which is in good accordance with the simulation results (from 942 to 1894 nm; the largest deviation between them is 1.8%). The strong dependence of resonance wavelength on the surrounding medium index is valuable for the sensing applications [3].

### IV. Dipole moment

The plasmonic resonance of the nanorod is excited by the incident light. Thus the conduction current and surface charges will be strongly dependent on the light electric field $E_0$. In the subwavelength region ($2r_0 << l << \lambda$), such a relationship may be bridged approximately with the circuit theory. Nonetheless, the non-uniform current distribution along the rod axis hinders the direct application of the circuit equation. To address this question, an equivalent method will be used in the following. Another key point is that the radiative resistance of the nanorod, which is not present in the conventional circuit theory, will be counted in our equivalent circuit equation.

Considering the circuit parameters introduced above, an effective induced electromotive force $-L_F dI_c/dt$ (where $I_c = J_0 S e^{-i\omega t}$ is the complex current at the center of nanorod) and an effective electric potential difference $-Q/C$ will be exerted on the nanorod. The equivalent circuit equation, which determines the central current $I_c$ of the nanorod, may be written as

$$\xi_{eff} - L_F \frac{dI_c}{dt} - \frac{Q}{C} = I_c (R_{rod} + R_{rad}), \tag{14}$$

where $\xi_{eff}$ represents the effective electromotive force contributed by the light electric field, $R_{rod}$ is the effective Ohmic resistance (or impedance) of the nanorod, and $R_{rad}$ is the effective resistance resulting from the radiation loss. The radiation of light will consume the energy of oscillating free electrons, thus an additional damping force is applied to the electrons, leading to an effective radiative resistance. Equation (14) can also be derived with the energy conservation (the energy of incident light will be converted to the kinetic energy of free electrons, the electric- and magnetic-field energy in the near field, the Ohmic heat due to the collision of free electrons, and the radiation energy in the far field).

The effective electromotive force $\xi_{eff}$ will transfer the energy from the incident light to the nanorod with an average power $P_{in} = (1/2)\operatorname{Re}(I_c^* \xi_{eff})$ (here * represents the complex conjugation). The input power can be obtained by performing the integral over the whole nanorod:

$$P_{in} \approx \frac{1}{2}\operatorname{Re}\int_{-l/2}^{l/2}[J(z)S]^* E_0 dz = \frac{1}{2}\operatorname{Re}(I_c^* E_0 \frac{2l}{\pi}). \tag{15}$$

Thus, the effective electromotive force driven by the light field reads

$$\xi_{eff} = E_0 \Delta z = (2/\pi) E_0 l. \tag{16}$$

Here, $\Delta z = 2l/\pi$ is the effective length, which is also the central distance of positive and negative charges $\pm Q$.

To calculate the effective Ohmic resistance $R_{rod}$, one notes that the power of effective $R_{rod}$ should be equal to that of the real nanorod. That means

$$\frac{1}{2}|I_c|^2 R_{rod} = \frac{1}{2}\int |J(z)S|^2 dR, \tag{17}$$

where $dR = dz/\sigma S$. By using the Eq. (1) and $\sigma = \sigma_0/(1-i\omega\tau)$, the $R_{rod}$ can be derived as

$$R_{rod} = \frac{1}{\sigma}\frac{l}{2S} = R_{Ohm} - i\omega L_K. \tag{18}$$

Here, $R_{Ohm} = l/2\sigma_0 S = \gamma L_K$ is the dc Ohmic resistance, $L_K = \mu_0 l/2k_p^2 S$ (where $k_p = \omega_p/c = 1/\delta$) is exactly the kinetic inductance of the nanorod and associated with the kinetic energy of free electrons. Thus, in the optical frequency range, the nanorod behaves like the series connection of a pure resistance and an inductor.

The effective radiative resistance $R_{rad}$ can be determined with the radiation energy. That is, the power consumed in $R_{rad}$ equals the radiation power of the real nanorod. According to the Eq. (6), the dipole moment of the nanorod is $p = (2/\pi)Ql$ and thus the radiation power is

$$P_{rad} = \frac{1}{2}|I_c|^2 R_{rad} = \frac{\mu_0 n_d}{12\pi c}\omega^4 |p|^2. \tag{19}$$

The central current and surface charges satisfy the relationship $I_c = dQ/dt = -i\omega Q$. Consequently, the effective radiative resistance is expressed as

$$R_{rad} = \frac{2\mu_0 n_d}{3\pi^3 c}\omega^2 l^2 = \frac{8}{3\pi}(\frac{n_d l}{\lambda})^2 Z, \tag{20}$$

where $Z = \sqrt{\mu_0/\varepsilon_0\varepsilon_d}$ is the impedance of the surrounding medium.

By substituting $\xi_{eff}$, $R_{rod}$ and $R_{rad}$ into the Eq. (14), the relationship between the central oscillating current $I_c$ and light electric field $E_0$ can be established. Accordingly, the dipole moment induced in the nanorod is derived as

$$p = \frac{\varepsilon_0 A}{\omega_0^2 - \omega^2 - i\eta\omega}E_0, \tag{21}$$

where

$$\begin{aligned}A &= \frac{(2l/\pi)^2}{\varepsilon_0(L_K + L_F)}, \\ \eta &= \frac{R_{Ohm} + R_{rad}}{L_K + L_F}.\end{aligned} \tag{22}$$

Hence, the induced dipole moment is proportional to the external light electric field $E_0$ and resonant at the fundamental frequency $\omega_0$ with a bandwidth $\eta$ (the quality factor is $Q = \omega_0/\eta$). Especially, the resonance bandwidth $\eta$ is proportional to the total resistance ($R_{Ohm} + R_{rad}$) and inversely proportional to the total inductance ($L_K + L_F$). If both the radiative resistance and Faraday inductance are ignored, $\eta$ will be degenerated exactly to $\gamma$, i.e., the collision frequency of the free electrons. However, the presence of the radiative resistance and Faraday inductance will modify the width of resonance (the radiative resistance tends to increase, while the Faraday inductance tends to decrease the bandwidth; generally, a spectral broadening occurs for the fundamental resonance). This point will be confirmed by the simulations.

### V. Extinction spectrum

*5.1 Scattering, absorption, and extinction spectra*

The plasmonic resonance of the nanorod is accompanied by a strong scattering and absorption of light. To calculate the scattering, absorption, and extinction cross sections, we define the polarizability of nanorod as $\chi = p/\varepsilon_0 E_0$. By normalizing the radiation power $P_{rad}$ of the nanorod with the intensity of incident light $I_l = \varepsilon_0 c n_d E_0^2 / 2$, the scattering cross section can be expressed as $C_{sca} = k_0^4 |\chi|^2 / 6\pi$, where $k_0$ is the wavevector in the free space. Moreover, according to the Eq. (15), the power consumed in the nanorod can be rewritten as $P_{in} = \mathrm{Re}(\dot{p} E_0^*)/2$. Actually, this power includes both the Ohmic and radiation loss (as the radiation resistance has been taken into account in the circuit equation). Thus, by normalizing $P_{in}$ with $I_l$, the extinction cross section is obtained as $C_{ext} = (k_0/n_d) \mathrm{Im}\,\chi$. The difference between $C_{ext}$ and $C_{sca}$ gives rise to the absorption cross section: $C_{abs} = C_{ext} - C_{sca}$ (which can also be obtained with $C_{abs} = |I_c|^2 R_{Ohm}/2I_l$).

With the polarizability $\chi$ deduced from the Eq. (21), one obtains

$$C_{sca} = \frac{A^2}{6\pi c^4} \frac{\omega^4}{(\omega_0^2 - \omega^2)^2 + \eta^2 \omega^2},$$

$$C_{ext} = \frac{A\eta}{n_d c} \frac{\omega^2}{(\omega_0^2 - \omega^2)^2 + \eta^2 \omega^2}. \tag{23}$$

One can also obtain with the Eqs. (23) that $C_{sca}/C_{ext} = R_{rad}/(R_{Ohm} + R_{rad})$. At the fundamental plasmonic resonance of the nanorod, the maximum of the scattering, absorption, and extinction cross sections can be derived, respectively, as

$$C_{sca}^m = (\frac{2l}{\pi})^2 \frac{ZR_{rad}}{(R_{Ohm} + R_{rad})^2},$$

$$C_{abs}^m = (\frac{2l}{\pi})^2 \frac{ZR_{Ohm}}{(R_{Ohm} + R_{rad})^2}, \tag{24}$$

$$C_{ext}^m = (\frac{2l}{\pi})^2 \frac{Z}{R_{Ohm} + R_{rad}}.$$

Then, the ratio $Z/(R_{Ohm} + R_{rad})$ measures the maximal ability of energy transfer from the incident light to the nanorod. Note that the spectral width $\eta$ is proportional to the total resistance, i.e., $\eta \propto R_{Ohm} + R_{rad}$ [see Eqs. (22)]. Thus, the larger the total resistance, the lower the peak of extinction spectrum and the wider the spectral width. In addition, Eqs. (24) show that $R_{rad}/(R_{Ohm} + R_{rad})$ governs the radiation of energy from the nanorod to the surrounding space and $R_{Ohm}/(R_{Ohm} + R_{rad})$ dominates the fraction of energy dissipated in the Ohmic loss. The results provide a useful guideline for the optimization of light scattering or absorption of the metallic nanorods.

*5.2 Dependence of extinction spectrum on the parameters*

To grasp the detailed characteristics of the scattering, absorption, and extinction spectra, we take the nanorod with a length $l$=160 nm and radius $r_0$=10 nm (in vacuum) as an example. Figure 3(a) presents the calculation results by using the Eqs. (23), with the resonance frequency $\omega_0$ determined by the Eq. (13). One can see that there is a prominent peak, locating at the resonance wavelength $\lambda_0 = 934$ nm, in each spectrum. Thus, the fundamental plasmonic resonance of the nanorod is accompanied by the strong scattering, absorption, and extinction of light. The full-width at half-maximum (FWHM) of the resonance peak is $\Delta\lambda = 39$ nm, thus corresponding to a Q factor of 23.9. The frequency range of the FWHM can be determined by $\Delta\omega = 2\pi c\,(\Delta\lambda/\lambda_0^2)$ or $\eta = (R_{Ohm} + R_{rad})/(L_K + L_F)$. Using the known parameters, we have $L_K$=0.155 pH, $L_F$=0.048 pH, $R_{Ohm} = \gamma L_K = 7.74\,\Omega$, and $R_{rad} = 9.38\,\Omega$. Hence, the spectral width is calculated as $\Delta\omega = \eta = 8.44\times10^{13}$ Hz, which is obviously larger than the employed electron collision frequency $\gamma = 5\times10^{13}$ Hz. In addition, the maximal scattering, absorption, and extinction cross sections can be obtained from the spectra [or with the Eqs. (24)] as 0.125, 0.103, and 0.228 μm², respectively. The maximal extinction efficiency $C_{ext}^m/2r_0 l$, i.e., the maximal extinction cross section normalized by the projection area of the nanorod (on the plane perpendicular to the light propagation direction), attains a large value of 71.

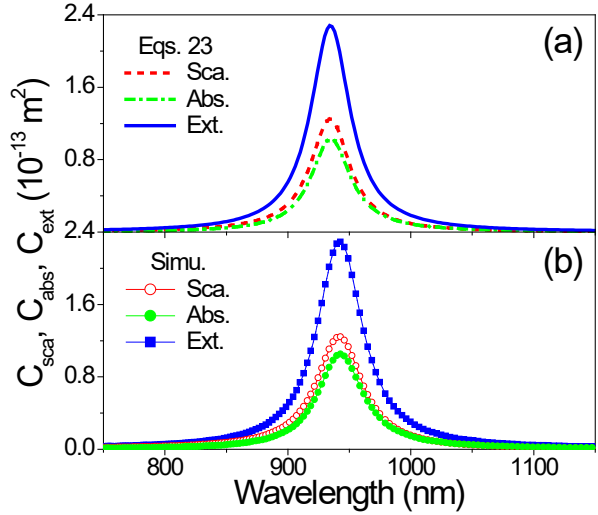

*Fig. 3 Scattering, absorption, and extinction spectra of the nanorod obtained with the analytical formulas (a) and numerical simulations (b). Here, the rod length is l=160 nm, the radius is $r_0$=10 nm, and the surrounding medium is vacuum.*

As a comparison, Fig. 3(b) presents the simulated scattering, absorption, and extinction spectra. The peaks of the spectra locate at the wavelength $\lambda_0$=942 nm with a FWHM of $\Delta\lambda = 44$ nm and a Q factor of 21.4. The frequency range of FWHM is $\Delta\omega = 9.34\times10^{13}$ Hz, which is also larger than $\gamma$ and close to the analytical value $\eta = 8.44 \times 10^{13}$ $Hz$. This confirms the role of the radiative resistance and Faraday inductance. Furthermore, the maximal scattering, absorption, and extinction cross sections given by the simulation are 0.124, 0.105, and 0.229 μm², respectively, which agree well with the analytical results (0.125, 0.103, and 0.228 μm²).

By varying the structural and material parameters, we investigated the evolution of extinction spectrum of the nanorod. The results are shown in Fig. 4, where the left column (a, c, e) represents the analytical results and the right column (b, d, f) the numerical simulations. Figures 4(a, b) show the variation of extinction spectrum with the increase of the aspect ratio κ, where $r_0$=10 nm and $n_d$=1. According to the analytical results, when κ varies from 5 to 7.5, 10, 12.5, and 15, the maximal extinction cross section $C_{ext}^m$ increases almost linearly from 0.131 to 0.211, 0.300, 0.397, and 0.500 μm² (along with the redshift of the resonance peak). The results are in agreement with the simulations, where $C_{ext}^m$ grows from 0.119 to 0.209, 0.306, 0.411, and 0.519 μm². Equation (24) suggests that $C_{ext}^m \propto l^2 Z /(R_{Ohm}+R_{rad})$, where $Z \propto 1/n_d$, $R_{Ohm} \propto l/r_0^2$, and $R_{rad} \propto (n_d l/\lambda_0)^2 Z$. In present case, the resonance wavelength is nearly proportional to the aspect ratio or rod length [see Fig. 2(a)], thus $R_{rad}$ is almost a constant at the resonance. Consequently, $C_{ext}^m \propto l^2/(l+c_1)$ (where $c_1$ is a constant), which varies almost linearly with $l$ (or κ) when $l$ is larger enough. Moreover, the extinction energy is separated into two parts, one is the scattering with the fraction $R_{rad}/(R_{Ohm}+R_{rad}) \propto 1/(l+c_1)$ and the other is the absorption with the fraction $R_{Ohm}/(R_{Ohm}+R_{rad}) \propto l/(l+c_1)$. The latter acquires a further enhancement, indicating that the longer nanorod (with the fixed radius) is more helpful for the light absorption and the photo-thermal effect.

Figures 4(c, d) exhibit the extinction spectra for different rod radius, where $n_d$=1 and the aspect ratio is fixed as κ=8. When the radius varies from 5 to 10, 15, and 20 nm, analytically $C_{ext}^m$ increases from 0.053 to 0.228, 0.411, and 0.608 μm²; and numerically, $C_{ext}^m$ grows from 0.053 to 0.229, 0.399, and 0.569 μm². The larger deviation of $C_{ext}^m$ (~7%) for $r_0$=20 nm is mainly caused by the deviation of resonance wavelength (~4.5%). If we use the numerical resonance wavelength (1168 nm) rather than the analytical value (1221 nm) to determine the radiative resistance $R_{rad}$, Eq. (24) will give $C_{ext}^m = 0.561$ μm², which is much closer to the simulated value 0.569 μm². When the aspect ratio is fixed, we obtain $C_{ext}^m \propto 1/(V^{-1} + c_2 \lambda_0^{-2})$, where $V$ is the volume of nanorod and $c_2$ is a constant. Thus, a larger nanorod with a longer resonance wavelength is beneficial for the extinction of light. Moreover, the fraction of scattering light is proportional to $1/(c_2 + \lambda_0^2/V)$, which can be further enhanced. On the contrary, a thinner nanorod with the fixed aspect ratio will own a larger fraction of light absorption.

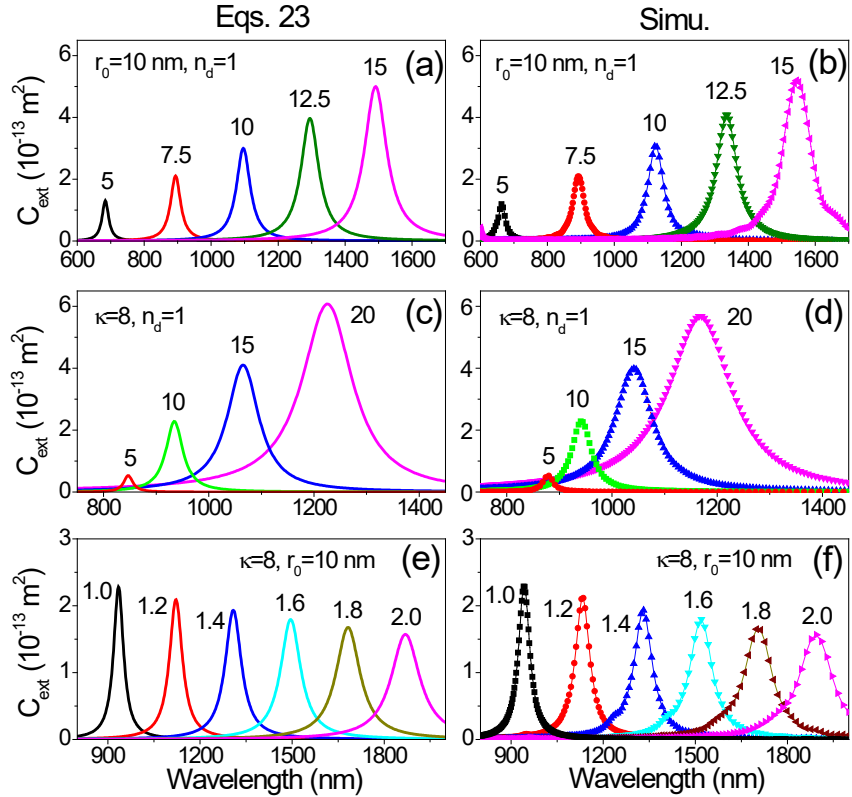

*Fig. 4 Extinction spectra of the nanorod: (a, b) $r_0$=10 nm, $n_d$=1, the aspect ratio is set as κ=5, 7.5, 10, 12.5, and 15, respectively; (c, d) κ=8, $n_d$=1, the rod radius is set as $r_0$=5, 10, 15, and 20 nm, respectively; (e, f) κ=8, $r_0$=10 nm, the refractive index of*

*the surrounding medium is set, respectively, as $n_d$=1.0, 1.2, 1.4, 1.6, 1.8, and 2.0. The left column (a, c, e) represents the analytical results obtained with the Eqs. (23) and the right one (b, d, f) the simulation results.*

Additionally, when the rod sizes are fixed and the refractive index of the surrounding medium is increased, $C_{ext}^m$ will be reduced along with the redshift of the resonance peak. The analytical results in Fig. 4(e) show that, when $n_d$ changes from 1.0 to 2.0 with an interval of 0.2 ($\kappa$=8, $r_0$=10 nm), $C_{ext}^m$ decreases gradually from 0.228 to 0.209, 0.193, 0.180, 0.168, and 0.157 μm². Again, the results are in good accordance with the simulation results in Fig. 4(f), where $C_{ext}^m$ reduces from 0.229 to 0.211, 0.193, 0.180, 0.165, and 0.157 μm². In this case, the radiative resistance $R_{rad} \propto (n_d l/\lambda_0)^2 Z$ will only vary with the impedance $Z$ of the surrounding medium, as the resonance wavelength is proportional to the medium index $n_d$ [see the Eq. (13) or Fig. 2(c)]. Hence, $C_{ext}^m \propto 1/(n_d + c_3)$ ($c_3$ is a constant), which decreases with the increase of medium index. The fraction of light scattering is also proportional to $1/(n_d + c_3)$ and degenerates simultaneously.

## VI. Field enhancement

The fundamental plasmonic resonance of the nanorod can induce a strong field enhancement near the metal surface [29]. On the surface of the two semispherical ends, surface charges are accumulated with a maximal density, thus leading to a maximal electric field $E_{end} = \rho_0/\varepsilon_0\varepsilon_d$. With the use of Eqs. (4) and (21), we obtain the frequency-dependent electric-field enhancement ratio as

$$\left|\frac{E_{end}}{E_0}\right| = \frac{\pi A}{4\varepsilon_d r_0 l^2} \frac{1}{\sqrt{(\omega_0^2 - \omega^2)^2 + \eta^2\omega^2}}. \tag{25}$$

At the plasmonic resonance, the maximal enhancement ratio becomes

$$\left|\frac{E_{end}}{E_0}\right|_{max} = \frac{\lambda_0}{2\pi^2 n_d r_0}\frac{Z}{R_{Ohm}+R_{rad}}. \tag{26}$$

On the other hand, the oscillating current induces a strong magnetic field, which is maximal at the waist ($r=r_0$, $z=0$) of the nanorod: $H_{mid}=J_0 r_0/2$. The magnetic-field enhancement ratio can be derived as

$$\left|\frac{H_{mid}}{H_0}\right| = \frac{A}{4 r_0 ln_d c}\frac{\omega}{\sqrt{(\omega_0^2-\omega^2)^2+\eta^2\omega^2}}. \tag{27}$$

The maximal enhancement ratio at the plasmonic resonance is

$$\left|\frac{H_{mid}}{H_0}\right|_{max} = \frac{2\kappa}{\pi^2}\frac{Z}{R_{Ohm}+R_{rad}}. \tag{28}$$

One can see that the maximal electric- and magnetic-field enhancement ratios are correlated with the peak value of extinction cross section [Eqs. (24)]. The larger the extinction cross section, the larger the energy stored in the system and magnitude of surface charges and conduction current, thus the larger the electric and magnetic field.

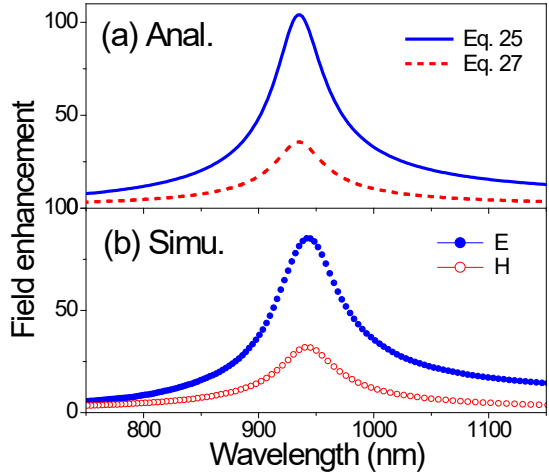

*Fig. 5 Enhancement of electric field (at the end of nanorod) and magnetic field (at the waist of the nanorod) as a function of wavelength: (a) Analytical and (b) simulated results. Here, the rod length is l=160 nm, the radius is $r_0$=10 nm, and the surrounding medium is vacuum.*

We take the nanorod with a length $l$=160 nm and radius $r_0$=10 nm (in vacuum) as an example. Figure 5(a) presents the analytical enhancement ratio of electric field (at the end of nanorod; the solid line) and magnetic field (at the waist of nanorod; the dash line) as a function of wavelength. The calculation shows that, near the wavelength 934 nm, the electric and magnetic fields are enhanced by 104.1 and 35.6 times, respectively. Also, by setting the probes at the end or waist of the nanorod, the enhancement ratios of electric field (the solid circles) and magnetic field (the open circles) have been simulated and plotted in Fig. 5(b). Near the wavelength 942 nm, the simulated electric field is enhanced by 85.4 times and magnetic field enhanced by 32.0 times, which are smaller but close to the analytical values [we found in the simulations that, by decreasing the mesh sizes of the nanorod, the numerical results of field enhancement will approach the analytical values gradually (here the mesh size was set as 0.2 nm)]. The small mode volume and strong near-field enhancement are important for boosting the light-matter interactions [30].

### VII. Conclusions

In summary, an analytical model has been developed to study the optical properties of a metallic nanorod, focusing on the longitudinal fundamental resonance mode. In this model, non-uniform distributions of conduction current and surface charges along the nanorod were considered. With the calculation of kinetic and electromagnetic energy, the circuit parameters of the nanorod have been determined. The plasmonic resonance wavelength predicted by theory agrees well with the simulations. Moreover, by introducing the radiative resistance to the circuit theory, the dipole moment, extinction spectrum, and near-field enhancement of the nanorod have been deduced analytically and confirmed numerically.

Our results are valuable for several reasons. Firstly, the results present a fast way to calculate the longitudinal fundamental resonance of the nanorod, such as the resonance wavelength, scattering, absorption, and extinction spectra, etc. In contrast, the pure numerical simulations rely heavily on the computer resources and are highly time consuming. Secondly, the results are useful for understanding the resonance effects, such as the spectral broadening of the fundamental resonance, the dependence of resonance wavelength and extinction spectrum on the geometrical or medium parameters, etc. Finally, the analytical formulas may provide a guideline for designing the nanorod sizes, thus enhancing/suppressing the light scattering at certain wavelength or maximizing the light absorption in the photo-thermal effect.

We emphasize that the present results such as the circuit parameters, resonance wavelength, and extinction spectrum etc. are based on the longitudinal fundamental current mode [the Eq. (1)]. Thus, the conclusion may not be applied (or directly applied) to the transverse plasmonic mode or high-order longitudinal modes of the nanorod [31]. Nonetheless, by gathering the specific distributions of conduction current and surface charges, it is still possible to study the resonance and scattering of the high-order longitudinal modes in a similar way.

**ACKNOWLEDGMENTS**

This work was supported by the National Natural Science Foundation of China (Grant No. 12174193).

**Appendix A: Kinetic inductance**

Compared with the electric- or magnetic-field energy of a system, the kinetic energy of free electrons in the metal can be neglected in the low-frequency band. But in the optical range, the situation changes significantly: the latter contributes a lot to the energy of system. The kinetic inductance originates from the kinetic energy or inertia of free electrons. For the nanorod studied here, the velocity of free electrons, $v = J(z)/(-ne)$, will vary in space. With the Eq. (1), the kinetic energy of all moving electrons, $U_K = \sum_i (1/2) m v_i^2$, can be calculated as

$$U_K \approx \int_{-l/2}^{l/2} \frac{1}{2} m \left[ \frac{J_0 \cos(\pi z/l) \cos \omega t}{-ne} \right]^2 nS dz$$
$$= \frac{1}{4} \frac{\mu_0 l}{k_p^2 S} I_0^2 \cos^2 \omega t. \qquad (A1)$$

Here, $I_0 = J_0 S$ and $k_p = \omega_p / c = 1/\delta$ ($\delta$ is the skin depth of the metal). In the calculation, the cross-sectional area of the nanorod is treated as a constant throughout the rod length $l$, which will overestimate the kinetic energy of electrons. But this effect is very weak, as the conduction current and electron velocity approach zero near the two ends of the nanorod.

For the spatially uniform current distribution $I$, the kinetic inductance $L_K$ can be simply introduced by $U_K = (1/2) L_K I^2$. Here, as the conduction current is varying along the rod axis, the real current $I$ is defined as the spatial maximum at the center of nanorod, i.e., $I_m = I_0 \cos \omega t$. Consequently, the kinetic inductance is written as

$$L_K = \frac{\mu_0 l}{2 k_p^2 S} = \frac{\mu_0 l}{2\pi} (\frac{\delta}{r_0})^2. \qquad (A2)$$

**Appendix B: Faraday inductance**

Compared with the kinetic inductance that is linked to the kinetic energy of free electrons, the Faraday inductance is correlated with the magnetic-field energy outside and inside the nanorod. According to the magnetic loop theorem, the magnetic field excited by the conduction current is $H_{in} = Jr/2$ in the nanorod and $H_{out} \approx JS/2\pi r$ outside but near the nanorod. The former corresponds to an internal magnetic-field energy of

$$U_M^{in} = \frac{\mu_0 l}{32\pi} I_0^2 \cos^2 \omega t. \tag{B1}$$

The external magnetic-field energy is difficult to determine accurately. As can be seen from Fig. 1(b), the magnetic field is mainly concentrated near the nanorod. Thus, as an approximation, we make a simple cutoff by considering the magnetic energy in a cylindrical region that has a length $l$, an inner radius $r_0$, and an outer radius $l$ (the outer radius is chosen in terms of the distribution length of the conduction current). The external magnetic energy is thus calculated as

$$U_M^{out} = \frac{\mu_0 l}{8\pi} (\ln 2\kappa) I_0^2 \cos^2 \omega t. \tag{B2}$$

With the Eqs. (B1) and (B2), the total magnetic-field energy within and outside the nanorod becomes

$$U_M = \frac{\mu_0 l}{8\pi} (\ln 2\kappa + \frac{1}{4}) I_0^2 \cos^2 \omega t. \tag{B3}$$

Therefore, by setting $U_M = (1/2) L_F I_m^2$, the Faraday inductance of the nanorod can be obtained as

$$L_F = \frac{\mu_0 l}{4\pi} (\ln 2\kappa + \frac{1}{4}). \tag{B4}$$

## Appendix C: Nanorod capacitance

The accumulation of free charges on the side and ends of the nanorod induces an enhanced electric field near the nanorod [see Fig. 1(c)]. For the two semispherical ends, the electric field excited outside (in the upper or lower space) can be obtained with the Gauss theorem as

$$E_r^{end} = \frac{i\rho_0 r_0^2}{\varepsilon_0 \varepsilon_d r^2} e^{-i\omega t}. \tag{C1}$$

Consequently, the electric-field energy stored near the two semispherical ends becomes

$$U_E^{end} = \frac{2\pi r_0^3}{\varepsilon_0 \varepsilon_d} \rho_0^2 \sin^2 \omega t. \tag{C2}$$

In the middle cylindrical region ($r \geq r_0$; $-0.5l + r_0 \leq z \leq 0.5l - r_0$), the electric field is governed by the surface charges distributed on the side of nanorod. Following the surface charge distribution [see Eq. (2)], one may suppose that the radial component of electric field has a similar format:

$$E_r^{side} = A_0 \sin(\pi z / l) e^{-i\omega t}, \tag{C3}$$

where $A_0$ is a coefficient to be determined. By using the Gauss theorem around the nanorod, the coefficient is written as $A_0 = i\rho_0 r_0 / \varepsilon_0 \varepsilon_d r$. Considering that the surface charge density and radial electric field grow from 0 to maximum when $z$ increases from 0 to $\pm l/2$, we also make a cutoff in the calculation of electric-field energy (i.e., the range of $r$ is chosen approximately from $r_0$ to $l/2$). Therefore, the electric-field energy associated with the free charges on the side of nanorod is

$$U_E^{side} = \frac{\pi r_0^3}{\varepsilon_0 \varepsilon_d} (\kappa - 2)(\ln \kappa) \rho_0^2 \sin^2 \omega t. \tag{C4}$$

Besides the above electric-field energy of the isolated free charges, there is an interaction between the positive and negative free charges on the nanorod, thus corresponding to a negative electric potential energy. This interaction energy can be counted simply by using the potential energy of a pair of positive and negative point charges ($\pm Q$) separated by a distance $\Delta z = 2l/\pi$, i.e., $U_E^p = -(\operatorname{Re} Q)^2 / (4\pi \varepsilon_0 \varepsilon_d \Delta z)$. With the use of Eq. (4), we obtain

$$U_E^p = -\frac{\kappa r_0^3}{\varepsilon_0 \varepsilon_d} \rho_0^2 \sin^2 \omega t. \tag{C5}$$

According to the above results [Eqs. (C2), (C4), and (C5)], the total electric energy stored in the plasmonic system can be calculated as

$$U_E = \frac{(2\pi - \kappa) + \pi(\kappa - 2)\ln \kappa}{8\varepsilon_0 \varepsilon_d \kappa l} Q_0^2 \sin^2 \omega t. \tag{C6}$$

By setting $U_E = Q_m^2 / 2C$ (where $Q_m = Q_0 \sin \omega t$), the capacitance is determined as

$$C = \frac{4\varepsilon_0 \varepsilon_d \kappa l}{(2\pi - \kappa) + \pi(\kappa - 2)\ln \kappa}. \tag{C7}$$